\documentclass[useAMS,usenatbib]{mn2e}
\usepackage{graphicx}

\title[North-south asymmetry in solar activity:  predicting 
the amplitude of the next solar cycle]
{North-south asymmetry in solar activity:  predicting 
the amplitude of the next solar cycle}

\author[J. Javaraiah]{J. Javaraiah\thanks{E-mail: jj@astro.ucla.edu}
\thanks{Permanent address :
 Indian Institute of Astrophysics, Bangalore-560 034, India}\\
Department of Physics and Astronomy, 430 Portola Plaza,
University of California, Los Angeles, CA 90095, U. S. A.}
\begin{document}

\maketitle

\label{fristpage}

\begin{abstract}
Using 
  Greenwich and SOON sunspot group data during 
the period 1874\,--\,2005,  
we find   that the sums of the areas of the sunspot groups in    
$0^\circ$\,--\,$10^\circ$ 
latitude-interval of the
 Sun's northern hemisphere and 
in the time-interval, minus 1.35 year to plus 2.15 year
 from the time of the preceding minimum--and in the same 
latitude interval of the southern hemisphere 
but  plus 1.0 year to plus 1.75 year from the time of the maximum--of
 a sunspot cycle are  well correlating  with the amplitude (maximum
 of the smoothed monthly sunspot number) of its immediate following cycle. 
Using this relationship it is possible to predict the  amplitude of
 a  sunspot cycle by about 9\,--\,13 years in advance.  
We predicted $74 \pm 10$  for the amplitude of the upcoming cycle~24. 
Variations in solar meridional flows  during solar cycles 
 and  9\,--\,16 year variations in   
 solar equatorial rotation may be responsible for the 
aforementioned relationship.
\end{abstract}


   \maketitle
    
\section{Introduction}
 The  prediction of the   
level of activity is important   because solar activity impact us 
in many ways~\citep{hwr,hw}. 
For example, solar flare activity  
cause geomagnetic storm that can
  cripple communication and damage power grids.
 There is also mounting evidence
that solar activity has an influence on terrestrial climate 
and space weather~\citep{roz,kmh,geo}. 
Many attempts have been made to predict the amplitude 
of a new sunspot cycle  
 by using old cycles data with a belief that  
 solar magnetic field 
persists for quite sometime~\citep{hwr}.  
 The existence of  a statistically significant 
difference between the levels of solar activity
 in the northern and the southern 
hemispheres    is 
shown by several 
statistical studies for most of the solar activity
 phenomena~\citep{garcia,car}. 
 The north-south asymmetry is unusually large 
during the Maunder minimum~\citep{sok}.  
The existence of  a few  periodicities in the north-south asymmetry of 
 solar activity is also shown~\citep{jg97,ksb}. 
In addition, there are  considerable north-south differences  
in the differential rotation rates  and the meridional motions of 
sunspots~\citep{ju}.     
 Helioseismology measurements also show
the existence of  north-south differences in the 
solar rotational and meridional flows~\citep{komm}. 
Therefore, north-south asymmetry in solar activity 
is an important physical solar property and it greatly helps 
for understanding  variations in  
the solar activity~\citep{sok,jg97,ksb}.  
In this letter we have used this  property of  
a  solar cycle to predict the amplitude of the upcoming solar
 cycle~24.

\section{Data analysis and Results}
We have used       
the Greenwich sunspot group data 
during the period 1874\,--\,1976, and 
the sunspot group data from the Solar Optical 
Observing Network (SOON)  of the US Air Force/US National Oceanic and 
Atmospheric Administration  during 1977 January 1\,--\,2005 September 30. 
We have 
 taken recently updated these data  from the NASA web-site of 
David Hathaway ({\tt http://solarscience.msfc.nasa.\break gov/greenwich.shtml}). 
These data include  the observation time (the date and the fraction 
of the day), 
the heliographic latitude and the longitude, central meridian distance ($CMD$), 
the corrected whole spot area (in mh), etc. for each day of the 
spot group observation
(130 mh  
 $\approx\ 10^{22}$ Mx). 
In the present analysis we have excluded the data corresponding to the 
$|CMD| > 75^\circ$ in  
any day of the spot group life-time. This precaution  
considerably  reduces the errors in the derived results due to the  
foreshortening effect. 
In case of SOON data, we increase area by a factor of 1.4. 
 David Hathaway found this correction is necessary to
 have a combined homogeneous 
 Greenwich and SOON
data (see aforementioned web-site of David Hathaway.)  
  We binned the daily data into  $10^\circ$  latitude 
intervals, in both the northern and the southern hemispheres,   
and determined  sum of the areas 
($AT$) 
of the spot groups in each $10^\circ$  
latitude interval, separately  for the rising and the declining phases of the
sunspot cycles 11\,--\,23. 
(It should be noted here that  cycle~23 is not yet complete.  
 The   data
are available for about 9 years of this cycle.   
 In case of cycle~11, the data  are available for the last 4 years.) 

We determined cross-correlations  between  
$AT$ and amplitude of cycle ($RM$).  
 We have taken the values of $RM$ (which is the largest smoothed monthly 
mean sunspot number),  and the epochs of maxima 
($TM$) and 
the preceding minima 
($Tm$) of 
 cycles  12\,--\,23   
from the  web-site, 
{\tt ftp://ftp.ngdc.noaa.gov/STP/SOLAR\_DATA/SUNSPOT\_NUMB\break ERS}. 
 Fig.~1 shows the  cross-correlation function, 
  $CCF(RM, AT)$, in different latitude intervals    
({\it a positive value of lag
 indicates that  
 $RM$ leads $AT$}).
 In this figure it can be seen that  except for $AT$  during the declining phases
of the cycles and in   
$0^\circ$\,--\,$10^\circ$
  latitudes intervals of 
both the northern and the southern hemispheres, 
for each of the remaining cases, 
 $viz.$  $AT$ in 
$10^\circ$\,--\,$20^\circ$ 
and 
$20^\circ$\,--\,$30^\circ$ 
latitude intervals  
 during the declining phases of the cycles and in all the 
latitude intervals during the rising phases of the cycles,    
 the corresponding
  $CCF(RM, AT)$   
has a weak peak  at $lag \ge 0$. 
This  suggests that in  all these latitude  intervals  
$AT$ and $RM$ variations  are approximately in the same phase   
or $RM$ leads  $AT$. 
In case of  $AT$  during the declining phases  of the cycles, 
   in  
$0^\circ$\,--\,$10^\circ$
  latitude interval
of the southern hemisphere the  $CCF(RM,AT)$ 
has a well defined peak (value 0.76) at $lag = -1$,  
suggesting that 
$AT$ leads that of  $RM$ by  about 5\,--\,10 years.  
In the same latitude interval of the  northern 
hemisphere the  $CCF (RM,AT)$ is found to be having a 
broad peak with two humps  
 (values 0.8 and 0.6) at  
 $lag = 0$ and $lag = -2$, 
  suggesting that  
$AT$ leads $RM$ by about 5\,--\,25 years. 
These  results  
indicate that $AT$ can be used to predict $RM$.

\begin{figure*}
\centering
\includegraphics [width=\textwidth]{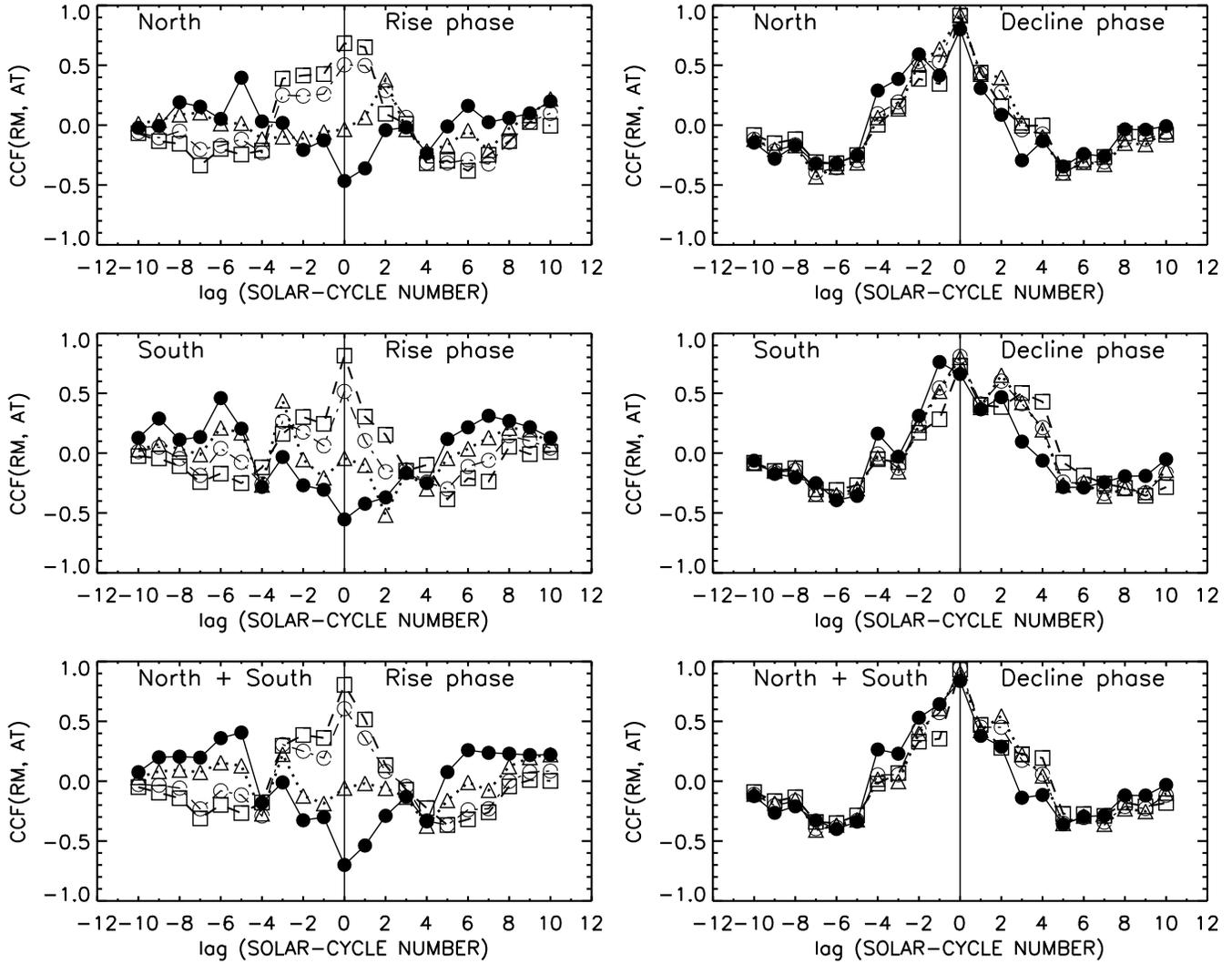}
\caption{Plots of the $CCF(RM,AT$) verses lag 
during the 
rising and declining phases of solar cycles~12\,--\,13.
A positive value of lag
 indicates  that  
 $RM$ leads $AT$.
The {\it filled circle-solid curve, triangle-dotted curve, square-dashed curve , and open circle-dash-dotted 
curve} represent $CCF(RM,AT$) in latitude intervals  $0^\circ-10^\circ$,
$10^\circ-20^\circ$, $20^\circ-30^\circ$  and in whole disk, respectively.}  
\end{figure*}

There exist a  number of  short-term periodicities, a few days to a few years,  
 in both the solar activity and the solar rotation~\citep{bs93,jjk,ksb}. 
Amplitude of such a periodicity largely varies during a solar cycle. 
 Therefore, there is a possibility that   
$AT$  in  
 $0^\circ$\,--\,$10^\circ$ 
latitude intervals of the northern and the southern hemispheres 
 during some  short intervals having  
strong correlations 
 with $RM$.   
 With this hypothesis  we determined 
the maximal values of
 correlations between $AT$   
of  cycle $n$
 and  $RM$ of cycle $n+1$ in the following way,     
where $n = 12,...,22$
is the cycle number: 
 First we  determined the values of $AT$ in the 
intervals which were  chosen arbitrarily  around 
the epochs of the maxima and the preceding minima of the cycles.  
The $AT$ determinations are repeated   
 by increasing or decreasing  
 the lengths of the intervals with a step of $\ge$ 0.05 year at a time. 
We  find that in 
$0^\circ$\,--\,$10^\circ$ 
latitude   interval of the 
southern hemisphere, the correlation is maximum,  
 coefficient of correlation
$r = 0.97$ (from eleven data points), in the short (0.75 year) 
 time-interval 
 just after 1-year after    
 the time of maximum of each of the cycles~12\,--\,23, 
 $TM^*$ : $TM + (1.0\ to\ 1.75)$ 
 ($i.e.$,  close to the time of
 the reversal of polarities of  
the polar magnetic fields). 
We also  find that in 
$0^\circ$\,--\,$10^\circ$
  latitude interval of the northern 
hemisphere $r=0.95$  
is maximum in 
the time-interval (3.5 year), 
 $Tm^*$ : $Tm + (-1.35\ to\ 2.15)$.
Both these correlations are statistically high significant with $>$ 99.99
 confidence
level (from {\it Student's t-test}), $i.e.$, the chance of getting
 these relations 
 from uncorrelated quantities is less than  0.01\%. 
Interestingly, the existence of 0.75 year periodicity is known  
 in solar activity~\citep{ksb}, and  
it may be a subharmonic of the well-known Rieger periodicity
 in solar flare activity~\citep{bs93}. 
The existence of  3.5 year periodicity in solar activity is 
also known and this   
 periodicity seems to be more pronounced in the north-south 
asymmetries of solar 
activity and surface rotation~\citep{jg97,ksb}.
 In Table~1 we have given the  values of  $AT$  
 during 
$Tm^*$ and $TM^*$. 
In the same table  we have also  given the values of    
 the amplitudes and the epochs 
of  maxima and minima of the sunspot cycles 12\,--\,23.

\begin{table*} 
\flushleft
{\scriptsize
   \caption{The maximum ($RM$) and the minimum 
($Rm$) amplitudes 
(the largest and the smallest  smoothed monthly mean sunspot numbers)
of the  solar-cycles 12\,--\,23 and   
the sum of the areas of spot groups  ($AT$, normalized by 1000)
 in the intervals 
$Tm^* = Tm + (-1.35\ to\ 2.15)$ and 
$TM^* = TM + (1.0\ to\ 1.75)$, 
where $TM$ and $Tm$ represent  the maximum
 and the preceding minimum epochs of the 
solar cycles, respectively.} 

\begin{tabular}{lcccccccccccc}
\hline
  \noalign{\smallskip}
Cycle && \multicolumn{2}{c}{Minimum} && \multicolumn{2}{c}{Maximum} && 
\multicolumn{2}{c}{Latitude Int.: $0^\circ - 10^\circ$ (north)} && \multicolumn{2}{c}{Latitude Int.: $0^\circ - 10^\circ$ (south)} \\
  \noalign{\smallskip}
\cline{1-1}
\cline{3-4}
\cline{6-7}
\cline{9-10}
\cline{12-13}
  \noalign{\smallskip}

$n$&&$Tm$&$Rm$&&$TM$&$RM$&&$Tm^*$&$AT$&&$TM^*$&$AT$\\

\hline
  \noalign{\smallskip}

12&&1878.9&2.2 &&1883.9& 74.6 &&1877.55\,--\,1881.05& 9.47&&1884.90\,--\,1885.65&42.11\\
13&&1889.6&5.0 &&1894.1& 87.9 &&1888.25\,--\,1891.75& 3.22&&1895.10\,--\,1895.85&32.64\\
14&&1901.7&2.6 &&1907.0& 64.2 &&1900.35\,--\,1903.85&12.98&&1908.00\,--\,1908.75&54.64\\
15&&1913.6&1.5 &&1917.6&105.4 &&1912.25\,--\,1915.75& 3.74&&1918.60\,--\,1919.35&34.58\\
16&&1923.6&5.6 &&1928.4& 78.1 &&1922.25\,--\,1925.75&33.96&&1929.40\,--\,1930.15&75.96\\
17&&1933.8&3.4 &&1937.4&119.2 &&1932.45\,--\,1935.95&29.96&&1938.40\,--\,1939.15&82.01\\
18&&1944.2&7.7 &&1947.5&151.8 &&1942.85\,--\,1946.35&69.35&&1948.50\,--\,1949.25&119.65\\
19&&1954.3&3.4 &&1957.9&201.3 &&1952.95\,--\,1956.45&15.23&&1958.90\,--\,1959.65&53.01\\
20&&1964.9&9.6 &&1968.9&110.6 &&1963.55\,--\,1967.05&50.31&&1969.90\,--\,1970.65&78.28\\
21&&1976.5&12.2&&1979.9&164.5 &&1975.15\,--\,1978.65&60.05&&1980.90\,--\,1981.65&83.53\\
22&&1986.8&12.3&&1989.6&158.5 &&1985.45\,--\,1988.95&29.85&&1990.60\,--\,1991.35&67.48\\
23$^{\mathrm{a}}$&&1996.4&8.0 &&2000.3&120.8 &&1995.05\,--\,1998.55&21.99&&2001.30\,--\,2002.05&33.58\\
\hline
  \noalign{\smallskip}

\end{tabular}

$^{\mathrm{a}}$ indicates the incompleteness of the current cycle~23.
}
\end{table*}

 We find the following   linear regressions fits between 
 $AT$ and $RM$ correspond
 to the correlations  above:
$$RM_{n+1} = (1.72 \pm 0.19) \times AT_n (Tm^*) + (74.0 \pm 7.0) , \eqno(1)$$
$$RM_{n+1} = (1.55 \pm 0.14) \times AT_n (TM^*) + (21.8 \pm 9.6) , \eqno(2)$$
\noindent where uncertainties in the coefficients are
 the formal 1-$\sigma$ (standard deviation) errors from the fit. 
In  equations~(1) and (2)
 the   slopes are 
    on 9$\sigma$ and 11$\sigma$  levels, respectively.  
That is, they are statistically high significant. 
 Therefore, the  relationship between  $AT_n$  and 
$RM_{n+1}$ is well described by these linear equations. 
It should be noted here that 
 always  
$Tm^*$ is  
associated with 
$0^\circ$\,--\,$10^\circ$
 latitude interval in the
 northern hemisphere, whereas  
$TM^*$ is associated  with 
$0^\circ$\,--\,$10^\circ$ 
 latitude 
interval of the southern hemisphere (for other combinations, $i.e.$,  
 $TM^*$  with  
$0^\circ$\,--\,$10^\circ$ 
interval 
of the northern hemisphere and 
$Tm^*$ with  
$0^\circ$\,--\,$10^\circ$
 interval 
of the southern 
hemisphere the values of $r$ found to be mere 0.11 and -0.24, respectively).

 Using  equations~(1) and (2)    
the amplitudes  of the upcoming sunspot cycles can be predicted
  by about 13 years 
and 9 years in advance, 
respectively.
The results of the least-square fits
are shown in Fig.~2(a). 
Fig.~2(b)  shows the 
correlation between the simulated  amplitudes ($PM$) [simulated using
 equations~(1) and (2)]  
 and the observed amplitudes ($RM$) of the cycles 
13\,--\,23. The  correlations 
between $PM$ and $RM$ and 
their levels of significance  are the same  
as those of  $AT_n$ and $RM_{n+1}$. 

\begin{figure*}
\centering
\includegraphics [width=\textwidth]{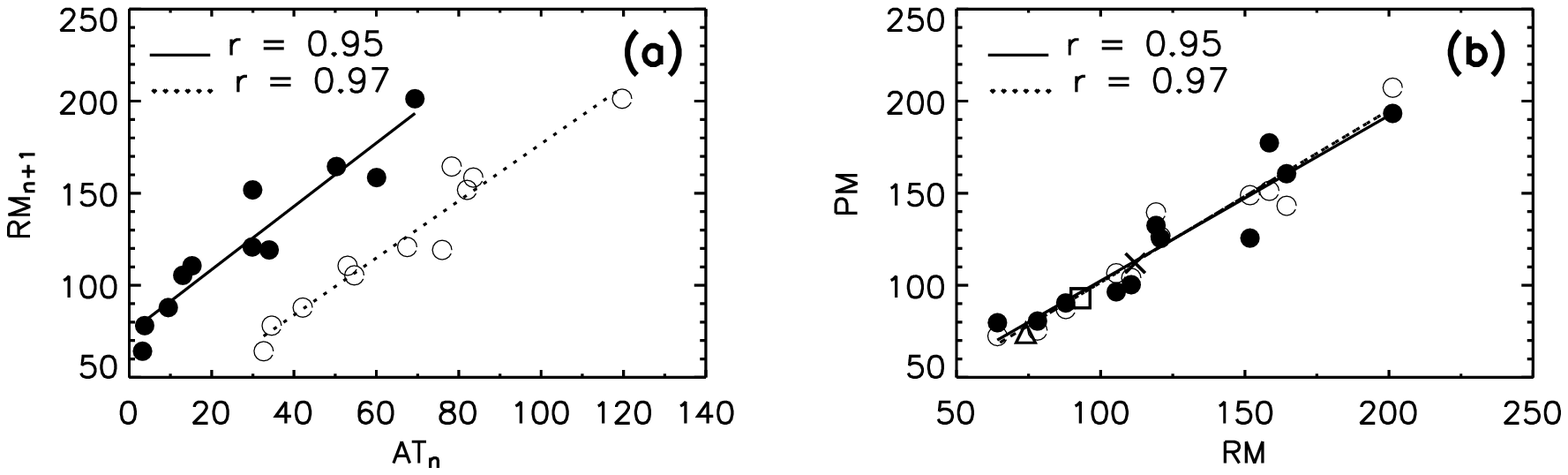}
\caption{Plots of the correlations (a) between 
the  $AT$ (for the values given in Table~1) during
 the intervals 
$Tm^*$ and $TM^*$ 
correspond to   
 cycle $n$
 and  $RM$ of the cycle $n+1$, and (b) between 
$RM$ and the simulated amplitude $PM$ of  cycle $n+1$, 
where $n = 12,...,22$,
is the cycle number.  
The straight lines represent the corresponding linear relationships.   The 
values of the correlation coefficient ($r$) are also given.
 The {\it filled circle} 
and the {\it solid line} correspond to the $AT$ during 
$Tm^*$ and the {\it open circle} and {\it dotted line} 
 correspond to the $AT$  during   
$TM^*$. The {\it cross} and {\it triangle} represent 
the values for  $RM$ of cycle~24  
obtained using $AT$ during $Tm^*$ and $TM^*$, 
respectively, and the {\it square} 
represents the corresponding mean value. We predict  
the value represented by the {\it triangle} for $RM$ of cycle~24.} 
\end{figure*}

Using 
 equation~(1) and (2) we obtained the values  
 $112 \pm 13$  and $74 \pm 10$, respectively,  for $RM$ of the  
  upcoming cycle~24 
(the uncertainty is 1$\sigma$ value).
The latter is  more statistically significant than  the former. 
Hence, by using equation~(2) the amplitude 
of a cycle can be predicted accurately 
by 9 years advance. 
  Therefore, we predict $74 \pm 10$ 
 for $RM$ of  cycle~24. This    
  is equal to the value predicted by~\cite{sva} (see Section~3).
The pattern of the mean cycle-to-cycle variation of
the  simulated  amplitudes ($PM$) obtained using 
equations (1) and (2) is found to be  
 slightly more strikingly
resemble with that of $RM$ ($r = 0.97$).
From this  we get $93 \pm 10$ for $RM$ of cycle~24.    
However, the difference between 
the   values obtained from equations~(1) and (2)  for cycle~24
 is significantly large.  
 The mean deviation is at $2 \sigma$ level. 
Hence, we do not suggest the mean value for $RM$ of cycle~24. 
Moreover, from equations~(1) and (2) we can get 
$RM_{n+1} \approx 2.1 \times AT_n (TM^*) -  0.6 
\times AT_n (Tm^*)$. [This  may be a more appropriate representation,  
because this is included both terms, $A_n (Tm^*)$ and $AT_n (TM^*)$.] 
 From this relation we get a much smaller value, 
  $57 \pm 13$, for the amplitude of cycle~24
($r = 0.95$). It  is somewhat  closer to the value obtained from 
equation~(2).
[The negative sign of the coefficient of $AT_n (Tm^*)$ in the aforementioned  
relation  can be attributed to the opposite  polarities  
of the magnetic fields    
 at $Tm^*$ and  $TM_{n+1}$ (in sunspot latitude belt).] 

Each of  the above derived values 
 for the amplitude of   
the upcoming cycle~24 is less than the $RM$ of cycle~23. This  
is consistent with the indication that the level of activity is 
now  at the declining phase of the current Gleissberg cycle~\citep{jbu}.  
From equations~(1) and (2) we can also get   
$AT_n (TM^*) \approx 1.11 \times AT_n (Tm^*) + 33.6$. [$r = 0.94$, between
 the simulated 
 and the observed $AT (TM^*)$. Note: the residual is quite large in case of cycle~23.] 
Hence, the magnetic field at $Tm^*$  may contribute to the field 
at  $TM_{n+1}$ both directly 
and through influencing the field at $TM^*$.
There is also a suggestion that when $AT_n (Tm^*)$  
 is zero the $AT_n(TM^*)$ is not always zero. 
 This might have happened during the late Maunder 
minimum, when sunspot activity is somewhat more pronounced 
in the southern hemisphere than in the northern hemisphere~\citep[see][]{sok}. 
[The current cycle~23 will be  ending soon. So, using equation~(1),
 or   using the aforementioned relationship between  
$AT_n (Tm^*$)   and $AT_n (TM^*$)  and 
equation~(2), an approximate prediction can be 
made for the amplitude of  cycle~25 in a 3 years time.]

\section{Discussion}
The  strength of the  preceding minimum is used   
to predict the strength of  the maximum of the same cycle. However,  
 it seems this methods works better after 1\,--\,2  year
 after the start of the cycle, $i.e.$, 
an accurate  prediction is possible 
only by about 3\,--\,4 years advance. 
The same is also true for    
the predictions based on geomagnetic indices 
   as precursor indicators~\citep{hwr}.

The  magnetic fields at the Sun's
  polar regions  are important ingredient for a dynamo model~\citep{ub04}. 
The polar field is maximum near sunspot minimum.
\cite{sch} have used,  for the first time,     
the strength of the  polar fields at the preceding minimum of a cycle 
as a precursor indicator to the strength of the following maximum. 
Recently, \cite{sva}  analyzed    
 the polar fields data during the  recent four solar cycles and predicted 
a small amplitude, $75 \pm 8$, for the  upcoming cycle~24.
Obviously from this method  the 
prediction  can be made   
only by about  5 years in advance. This method seems to be more  uncertain 
and could fail if used too early before the start 
of the cycle~\citep{sva}.

\cite{dg06},
 by   simulating  the surface 
magnetic flux using the guidelines of a dynamo model,  predicted 
 a  large amplitude, 150\,--\,180, for   cycle 24, $i.e.$, 
 a contradiction to the aforementioned prediction by  
\cite{sva}.         
This discrepancy implies that the
 dynamo processes are not yet 
 fully understood, making 
prediction more difficult~\citep{tob}.

Using the well known Gnevyshev-Ohl rule or G-O rule~\citep{gnev} 
 it is possible to predict only the amplitude of an 
 odd numbered cycle~\citep{wil}.   
  This is also not always possible  because 
occasionally (for example, recently by the cycles' pair 22,23) the G-O 
rule is violated. 
A major advantage of the 
$AT_n$\,--\,$RM_{n+1}$ 
relationships  above is that 
 using these   
the amplitudes of both  
 odd and   even numbered cycles can be predicted. 
In addition,  
 this new method seems to have a solid     
 physical basis.   
Interestingly, the   
$TM^*$ is 
  very  close to the epoch when the  
 polar-fields polarities reversals  take place~\citep{mak03} and  
 $Tm^*$ is close to  the epoch when  
the magnetic fields polarities reversals 
take place close to the equator, $i.e.$, at the beginning of a cycle and 
 continuing through 
the years of minimum~\citep{mak01}. 
This suggest that the    
$AT_n$\,--\,$RM_{n+1}$
 relationships 
are related to the 22-year solar magnetic cycle.
It should be noted here that  although sunspot activity is confined 
to  middle 
and low latitudes, it may be caused by the global modes of solar 
magnetic cycle~\citep{gjkv,juckett}.   

 Reconnection of the magnetic fields of opposite
 polarities
 is believed to be  
the basic mechanism of flare  activity.   
During $Tm^*$ 
the magnetic field structure seem to be largely quadrupole 
nature, which is probably favorable 
for X-class flares production~\citep{garcia}. 
The solar meridional flows transport angular momentum and magnetic field from 
pole to equator and vise-versa, in the convection zone.  
The motions of  spot groups  mimic 
the motions in the 
convection zone~\citep{jg97b,jjk}.
  The mean  meridional motion of sunspot groups 
is changing from pole-ward to equator-ward
rapidly in  
$0^\circ$\,--\,$10^\circ$ 
  latitude interval 
of  the northern hemisphere and gradually in   
the same latitude interval of southern hemisphere
  during 
$Tm^*$  and $TM^*$, 
respectively~\citep[see Fig.~2 in][]{ju}.
These results  indicate a participation  
 of  the meridional flows in the magnetic  
reconnection process and the reversals of the polarities of magnetic fields
 during 
$Tm^*$ and $TM^*$.   
The interceptions  of the pole-ward and     
 the equator-ward meridional flows
 may be responsible for the quadrupole nature of magnetic fields   
 during 
$Tm^*$.   
 It seems that during rising phases of the cycles the 
flare activity is strong   in the northern hemisphere
and weak in the southern hemisphere, 
and this is opposite during the declining phases of the cycles~\citep{garcia}. 
 During the rising phases of the cycles the mean meridional velocity of  
spot groups is equator-ward in the northern hemisphere and pole-ward 
in the southern hemisphere. During the declining phases of the cycles
 the velocity is pole-ward
in both hemispheres, but 
  the variation is steep in the  southern hemisphere, mainly   
in 
$20^\circ$\,--\,$30^\circ$ 
latitude interval~\citep{ju}. 
In view of  the above  inferences,   the 
 north-south asymmetry in solar flare activity may  be  
related to the 
north-south asymmetry in the meridional flows. 
 The corresponding  losses  in the  
magnetic flux in the northern and the southern hemispheres caused by 
the reconnection processes may have 
a contribution for the  north-south asymmetries in solar 
magnetic field and in sunspot-activity.  

 The lengths of the intervals from the beginnings of
 $Tm^*$ and $TM^*$ of    
a preceding cycle to  $TM$ 
 of its following cycle vary 14\,--\,19 years and 7\,--\,11 years,
 respectively. 
The corresponding mean values  are found to be 
16 years and 9.6 years, respectively. 
Similar periodicities exist in both the equatorial rotation rate
 and the latitude 
gradient term of the solar rotation determined from the sunspot group
 data~\citep{jg97,jj05,geo}.
Therefore, variations in the solar  meridional flows
 during solar cycles and   
    9\,--\,16 year variations in the solar 
equatorial rotation may be  
responsible for the  
$AT_n$\,--\,$RM_{n+1}$
 relationships above.

\section{Conclusions}
Using Greenwich and SOON sunspot group data during the period 1874\,--\,2005 
 we find that: 
\begin{enumerate}
 \item The sum of the areas ($AT$) of the spot groups in 
$0^\circ$\,--\,$10^\circ$ 
latitude interval of the Sun's
 northern hemisphere during the interval  
$Tm^*$ : $Tm + (- 1.35\ to\ 2.15)$
in  a  cycle  
is well correlated 
with the  amplitude ($RM$) of its following cycle, 
where
 $Tm$ is the time (in years) of the  preceding minimum of the 
preceding cycle,  

 \item The $AT$ of the spot groups in    
$0^\circ$\,--\,$10^\circ$ 
latitude interval of the southern hemisphere during the interval   
$TM^*$ : $TM + (1.0\ to\ 1.75)$ in 
  a cycle is also well correlated 
 with  $RM$ of its following cycle,
where $TM$
 is the time (in years) of the maximum of the preceding cycle.

\item Using `(i)'  and  `(ii)'  
it is possible to predict $RM$ of a cycle by about 13 years 
and 9 years
 advance, respectively. 

\item
We predicted $74 \pm 10$  for $RM$ of  cycle~24.  

\item  
Variations in solar  meridional flows during
 solar cycles and   9\,--\,16 year variations in  solar 
equatorial rotation  
 may be  responsible for the 
 relations `(i)' and `(ii)'. 
\end{enumerate}

\section*{acknowledgments}
I thank  
the  referee, Dr. Leif Svalgaard, for  
helpful comments and  suggestions, and Dr. David H. Hathaway 
for valuable information on the data.
I also  thank Professor Roger K. Ulrich and Dr. Luca Bertello
   for fruitful discussion, and I  
 acknowledge the funding  by NSF grant ATM-0236682.

\bsp
\label {lastpage}
\end{document}